\documentclass[onecolumn]{elsart}
\usepackage{epsfig}
\renewcommand{\nuc}[2]{\hbox{$^{#1}$#2}}
\begin{document}

\begin{frontmatter}

\title{In-beam $\gamma$-ray spectroscopy at the proton dripline: \nuc{23}{Al}}

\author[NSCL,MSUPA]{A. Gade}
\author[NSCL]{P. Adrich}
\author[NSCL]{D. Bazin}
\author[NSCL,MSUPA] {M. D. Bowen}
\author[NSCL,MSUPA] {B. A. Brown}
\author[NSCL,MSUPA]{C. M. Campbell}
\author[NSCL,MSUPA]{J. M. Cook}
\author[NSCL,MSUPA]{T. Glasmacher}
\author[URS]{K. Hosier}
\author[NSCL,MSUPA] {S. McDaniel}
\author[URS] {D. McGlinchery}
\author[NSCL]{A. Obertelli}
\author[URS]{L. A. Riley}
\author[NSCL,MSUPA] {K. Siwek}
\author[SUR]{J. A. Tostevin}
\author[NSCL]{D. Weisshaar}

\address[NSCL]{National Superconducting Cyclotron Laboratory, East
Lansing, MI 48824, USA}
\address[MSUPA]{Department of Physics \& Astronomy, Michigan State
  University, East Lansing, MI 48824, USA}
\address[URS] {Department of Physics and Astronomy, Ursinus College,
     Collegeville, PA 19426, USA}
\address[SUR] {Department of Physics, Faculty of Engineering and
      Physical Sciences, University of Surrey, Guildford, Surrey GU2 7XH,
      United Kingdom}
\begin{abstract}
We report on the first in-beam $\gamma$-ray spectroscopy of
\nuc{23}{Al} using two different reactions at
intermediate beam energies: inelastic
scattering off \nuc{9}{Be} and heavy-ion induced one-proton pickup,
\nuc{9}{Be}(\nuc{22}{Mg},\nuc{23}{Al}$+\gamma$)X, at
75.1~MeV/nucleon. A $\gamma$-ray transition at 1616(8)~keV --
exceeding the proton separation energy by 1494~keV -- was observed
in both reactions. From shell model and proton decay calculations we
argue that this $\gamma$-ray decay proceeds from the core-excited
$7/2^+$ state to the $5/2^+$ ground state of \nuc{23}{Al}. The
proposed nature of this state, $[\nuc{22}{Mg}(2^+_1) \otimes \pi
d_{5/2}]_{7/2^+}$, is consistent with the presence of a
$\gamma$-branch and with the population of this state in the two
reactions.
\end{abstract}


\end{frontmatter}

Since its discovery in 1969~\cite{cerny69}, the neutron-deficient
nucleus \nuc{23}{Al} has attracted much attention. \nuc{23}{Al} is
four neutrons removed from stable \nuc{27}{Al} and is the last
proton-bound, odd-mass aluminum isotope known to exist. The low
proton separation energy of $S_p=122(19)$~keV~\cite{audi03} made
\nuc{23}{Al} a candidate for a proton halo system. From
the measurement of an enhanced reaction cross section, \nuc{23}{Al}
was indeed proposed to have a proton-halo structure with a $J^{\pi}=
1/2^+$ assignment suggested for the spin and parity of the ground
state~\cite{cai02,zhang03}. However, a $\beta$NMR measurement
clearly showed that the spin and the parity of the \nuc{23}{Al}
ground state are $J^{\pi}=5/2^+$~\cite{ozawa06}, in agreement with
the mirror nucleus \nuc{23}{Ne}. Excited states of \nuc{23}{Al} have
been studied in \nuc{24}{Mg}(\nuc{7}{Li},\nuc{8}{He})\nuc{23}{Al}
reactions~\cite{wie88,cagg01}, in $\beta$-delayed proton decay 
\cite{blank97}, in Coulomb breakup~\cite{gomi05} and most recently
in $\nuc{22}{Mg}+p$ resonant proton scattering~\cite{he07}. None of
these prior experiments was sensitive to $\gamma$-ray transitions in
\nuc{23}{Al}. In the present paper we report on the first in-beam
$\gamma$-ray spectroscopy study of \nuc{23}{Al}. Two complementary
reactions were used to excite \nuc{23}{Al}: inelastic scattering off
a \nuc{9}{Be} target and the heavy-ion induced one-proton pickup reaction,
\nuc{9}{Be}(\nuc{22}{Mg},\nuc{23}{Al}$+\gamma$)X.

Both measurements were performed with an exotic cocktail beam
composed of 32\% \nuc{22}{Mg} and 3\% \nuc{23}{Al}. This secondary
beam was produced in-flight by fragmentation of a 150~MeV/nucleon
\nuc{36}{Ar} primary beam delivered by the Coupled Cyclotron
Facility at NSCL on the campus of Michigan State University. The
primary \nuc{9}{Be} production target (893~mg/cm$^2$ thick) was
located at the mid-acceptance target position of the A1900 fragment
separator~\cite{a1900}; an achromatic aluminum wedge degrader of
300~mg/cm$^2$ thickness and momentum slits at the dispersive image
of the separator were used to select the secondary beam. As the available total
secondary beam intensity exceeded various rate limitations (timing detectors
before the reaction target and data acquisition), the slits were restricted to
$\Delta p/p$=0.5\% total momentum acceptance for the secondary beam resulting in
the very clean particle identification discussed later.

The \nuc{9}{Be}(\nuc{23}{Al},\nuc{23}{Al}$+\gamma$)\nuc{9}{Be}
inelastic scattering and the \nuc{9}{Be}(\nuc{22}{Mg},\nuc{23}{Al}$
+\gamma$)X one-proton pickup reaction were induced by a
188(4)~mg/cm$^2$ \nuc{9}{Be} target placed at the target position of
the S800 spectrograph~\cite{s800}. The reaction target was
surrounded by SeGA~\cite{sega} in its ``classic'' configuration with
nine and seven detectors, respectively, at central angles of
$90^{\circ}$ and $37^{\circ}$ with respect to the beam axis. The
SeGA high-purity germanium detectors are 32-fold segmented and allow
for an event-by-event Doppler reconstruction of the $\gamma$-rays
emitted in-flight by the reacted nuclei. The emission angle that
enters the Doppler reconstruction is determined from the location of
the segment with the largest energy deposition relative to the
target.

Event-by-event particle identification was performed in all entrance
and exit channels with timing detectors before the reaction target
and with the focal-plane detection system~\cite{yurkon99} of the
S800 spectrograph. The time-of-flight difference measured between
the two plastic scintillators located before the reaction target
allowed for an unambiguous identification of all constituents of the
incoming cocktail beam (see Fig.~2 of~\cite{gade07a} and Fig.~1
of~\cite{gade08}). The reaction residues emerging from the
\nuc{9}{Be} target were identified via their time-of-flight measured
by plastic scintillators and their energy loss determined with the
S800 ion chamber. A software gate on the incoming beam then allowed
the selection of only those reaction residues induced by the
projectile of interest (see also refs.~\cite{gade07a,gade08}).

\begin{figure}[h]
\epsfxsize 7.8cm \epsfbox{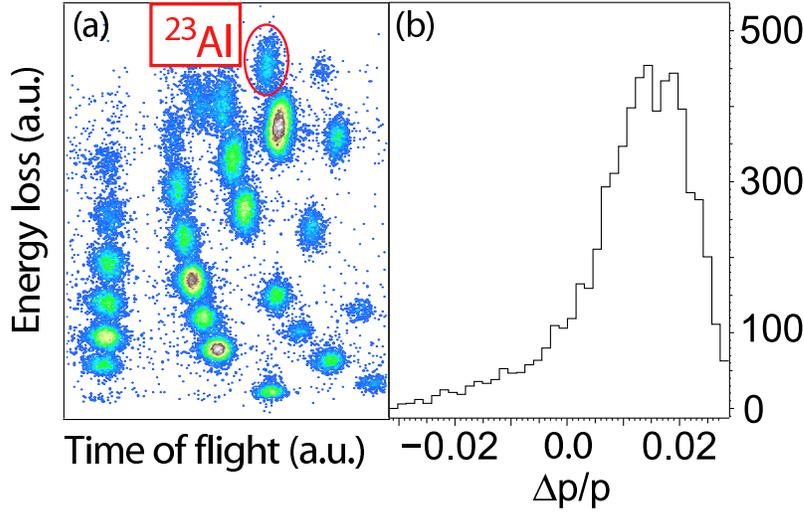}
\caption{\label{fig:pid_pick} Left: Identification spectrum --
energy loss vs. time of flight -- of \nuc{23}{Al} produced in
the one-proton pickup of \nuc{22}{Mg} projectiles from the \nuc{9}{Be}
target. Right: Parallel
momentum distribution of \nuc{23}{Al} (relative to $p=7.990$~GeV/c);
shown is the full momentum acceptance of the spectrograph.}
\end{figure}

\begin{figure}[h]
\epsfxsize 7.5cm \epsfbox{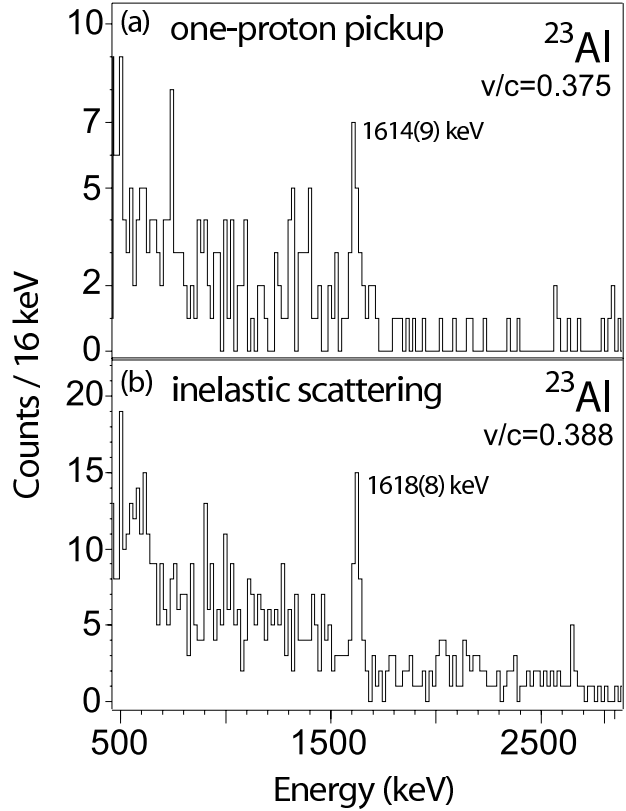}
\caption{\label{fig:spectrum} Doppler-reconstructed $\gamma$-ray
spectrum measured in coincidence with \nuc{23}{Al} reaction residues
produced in one-proton pickup (upper panel) and inelastic scattering
(lower panel). The energy uncertainty is dominated by the
uncertainty in the target position which is systematic and common
for both measurements.}
\end{figure}

Fig.~\ref{fig:pid_pick}(a) shows the particle
identification spectrum  -- energy loss vs. time-of-flight -- for
the spectrograph setting that was optimized on the one-proton pickup
channel. The particle identification spectrum only contains reaction
products from \nuc{22}{Mg}. The one-proton pickup residue
\nuc{23}{Al} can be clearly separated from the fragmentation
products that enter the S800 focal plane as well. Fig.~\ref{fig:spectrum}(a)
shows the event-by-event Doppler
reconstructed $\gamma$-ray spectrum in coincidence with \nuc{23}{Al}
produced in the one-proton pickup reaction. A photopeak at
1614(9)~keV  is clearly visible and marks a $\gamma$-ray transition
in \nuc{23}{Al}.

\begin{figure}[h]
\epsfxsize 7.8cm \epsfbox{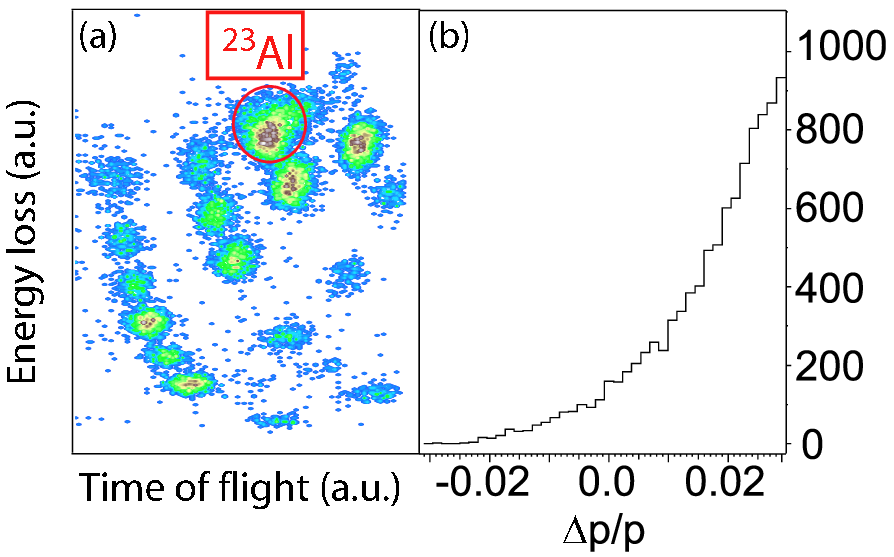}
\caption{\label{fig:pid_inel} Left: Identification spectrum --
energy loss vs. time of flight -- of the (in)elastically scattered
\nuc{23}{Al} nuclei. The spectrograph magnetic rigidity was centered
on the one-neutron removal products from \nuc{24}{Si}. Right: Parallel momentum
distribution of the scattered \nuc{23}{Al} (relative to
$p=8.339$~GeV/c). Due to the magnetic rigidity setting optimized on
reaction residues with one neutron less than the projectile, only
the low-momentum tail of scattered \nuc{23}{Al} enters the focal
plane. }
\end{figure}

 Fig.~\ref{fig:pid_inel}(a) shows the particle
identification of the inelastically scattered \nuc{23}{Al}. The
spectrograph setting was optimized on the one-neutron knockout
reaction \nuc{9}{Be}(\nuc{24}{Si},\nuc{23}{Si})X which is discussed
in~\cite{gade08}. For a single-neutron knockout setting in this mass
region, the magnetic rigidity of the spectrograph is more than 4\%
lower than for a setting centered on the projectiles passing through
the target. Thus only the low-momentum tail of the scattered
\nuc{23}{Al} projectiles can enter the spectrograph as displayed in
Fig.~\ref{fig:pid_inel}(b). In coincidence with
these scattered \nuc{23}{Al} nuclei, that must have been subject to a
significant momentum transfer, to be found in the outmost
low-momentum tail of the distribution, a $\gamma$-ray transition was
observed at 1618(8)~keV (Fig.~\ref{fig:spectrum}(b)).

The observed $\gamma$-ray energy is consistent for the two
measurements and implies an excited state at 1616(8)~keV in
\nuc{23}{Al}, 1494~keV above the proton separation energy. This
state has a significant $\gamma$-ray branch to the proton-bound
ground state. If this $\gamma$-ray transition were to originate from
an even higher excited state it would either populate a
proton-unbound excited state and could not have been observed here
as a coincidence with \nuc{23}{Al} is required or would feed an
excited state that also has a significant $\gamma$-ray decay branch
which then should have been detected as well.

\begin{figure}[h]
\epsfxsize 8.0cm \epsfbox{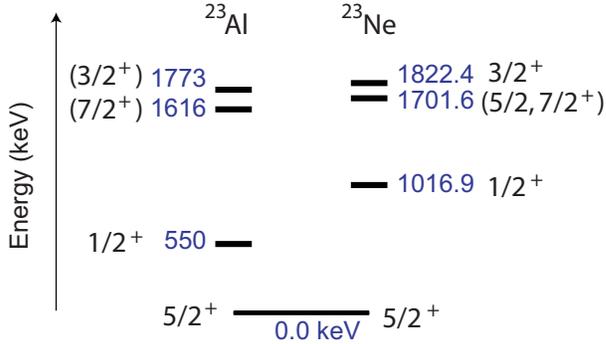} \caption{\label{fig:mirror}
Low lying level schemes of the mirror nuclei \nuc{23}{Al} and
\nuc{23}{Ne} below 2.3~MeV excitation energy. The large difference
in the energy of the $1/2^+$ state can be attributed to the
Thomas-Ehrman shift~\cite{thomas52} which is strongest for states
with orbital angular momentum $\ell=0$. The state at 1616(8)~keV in
\nuc{23}{Al} is reported for the first time.}
\end{figure}

All excited states of \nuc{23}{Al} are reported to decay to 100\% by
proton emission, including the first excited state, $E(1/2^+)
=550(20)$~keV, with an estimated branching of $\Gamma_p/\Gamma_{\gamma} \sim 10^{8}
$~\cite{cagg01,gomi05,datasheets}. Figure \ref{fig:mirror} compares the level
schemes of the mirror nuclei \nuc{23}{Al} and \nuc{23}{Ne} below
2.3~MeV excitation energy. We tentatively assign spin and parity
$J^{\pi}=(7/2^+)$ to the excited state at 1616(8)~keV observed in
the present work. There is a one-to-one correspondence for the
states below 2.3~MeV. The energy difference of the $1/2^+$ first
excited state can be explained by the Thomas-Ehrman
shift~\cite{thomas52} which influences $\ell=0$ orbits the most. In
the following we discuss the structure of the proposed $7/2^+$
state, in particular the occurrence of a significant $\gamma$-ray
branch, within the USD shell model and argue the consistency of this
assignment with the reaction mechanisms that led to its observation.

\begin{figure}[h]
\epsfxsize 7.8cm \epsfbox{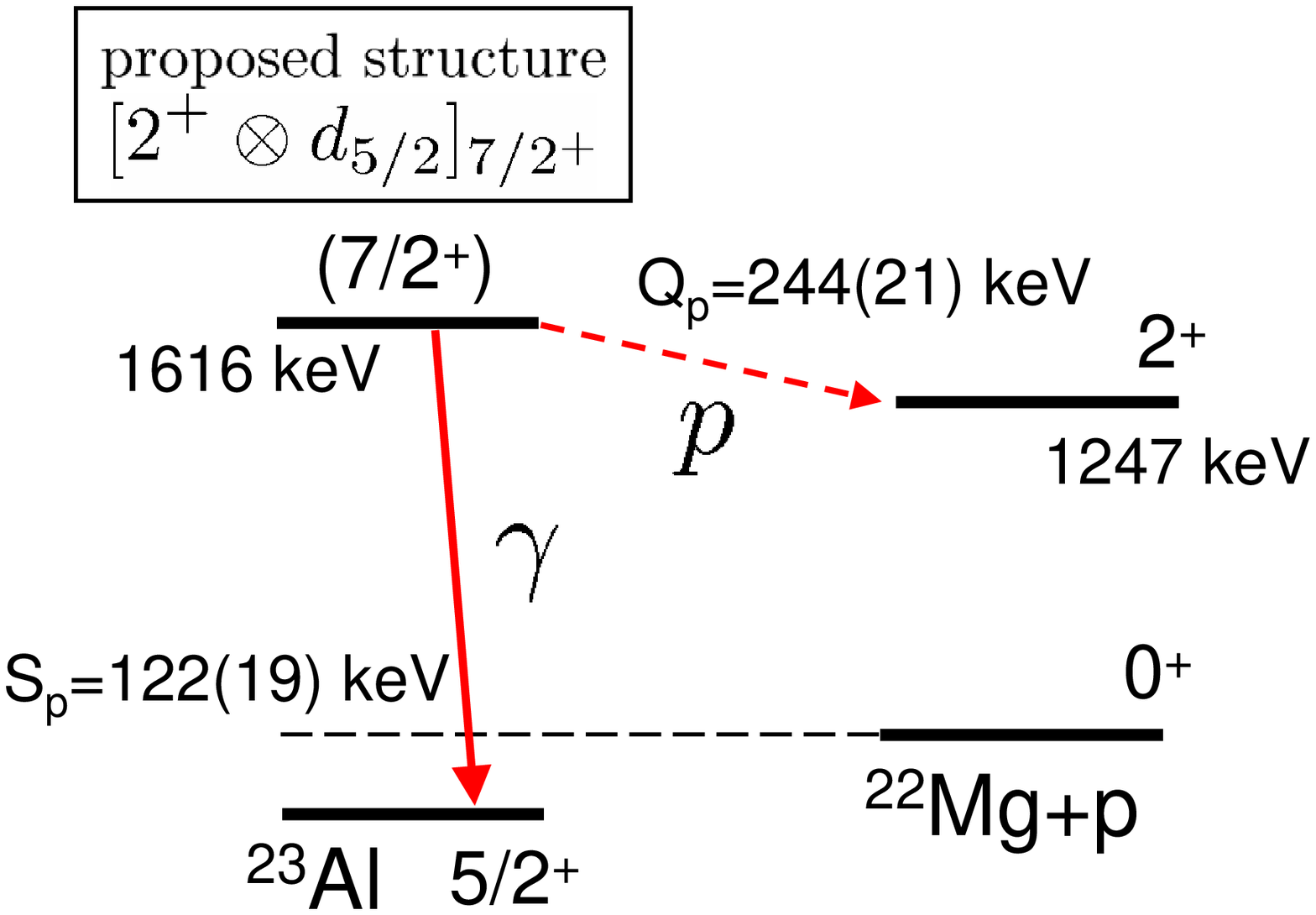} \caption{\label{fig:decay}
Proposed decay scheme of the $7/2^+$ core-coupled state at
1616(8)~keV in \nuc{23}{Al}. The state at 1616(8)~keV in
\nuc{23}{Al} was observed for the first time in the present
experiment.}
\end{figure}

Shell-model calculations for the energies, spectroscopic
factors and electromagnetic matrix elements were carried out with the
USDB effective interaction (results with USDA were similar) \cite{usd}.
The calculated half-life of the $7/2^+$ state, $T_{1/2}=23$~fs,
corresponds to a $\gamma$ width of $\Gamma_{\gamma}=0.020$~eV.
The spectroscopic factor for the $d_{5/2}$ orbit connecting the
\nuc{23}{Al}, $J^{\pi}=7/2^+$ and \nuc{22}{Mg}, $J^{\pi}=2^+$
levels is large indicating a dominance of the $[\nuc{22}{Mg}(2^+_1) \otimes
d_{5/2}]_{7/2^+}$ configuration in the \nuc{22}{Mg} wave function.
From the energetics summarized in Fig.
\ref{fig:decay}, it follows that the $Q$-value for this decay is
$Q_p=244(21)$~keV.


To quantify the proton decay of the $7/2^+$ state, the proton
scattering cross section was calculated for $\ell=2$ at $Q_p=244$~keV with a
Woods-Saxon potential and the resulting resonance width was used as
the single-particle proton decay width, $\Gamma^{sp}_p=0.0024$~eV.
With the value $S=0.46$ for the $7/2^+$ to $[\nuc{22}{Mg}(2^+_1)
\otimes d_{5/2}]$ spectroscopic factor, from the USDB shell-model
calculations, this yields a proton decay width of $\Gamma_p= S
\times \Gamma_p^{sp} = 0.0011$~eV for the decay of the $7/2^+$ state
of \nuc{23}{Al} to the first excited $2^+$ state of \nuc{22}{Mg}. In
conclusion, the proposed structure of the $7/2^+$ state, together
with the energetics of the proton decay (see Fig.~\ref{fig:decay}),
results in $\Gamma_{\gamma}/\Gamma_p \sim 20$ consistent with the
observation of the $\gamma$-decay of this state. We note that proton
detection was not possible with our experimental setup. Furthermore,
\nuc{22}{Mg} residues populated by the proton decays of excited
states of \nuc{23}{Al} could not be distinguished from \nuc{22}{Mg}
produced by the fragmentation of the \nuc{23}{Al} projectiles, for
example. Proton decay to the \nuc{22}{Mg} ground state could
proceed by $\ell=4$. The single-particle proton decay width
from the potential scattering calculations for
$\ell=4$ with $Q_p=1.49$~MeV is 10~eV. If we assume that the
proton width is approximately less
than or equal to the gamma width then $S \leq 0.002$ for the $g_{7/2}$
spectroscopic factor. Thus, the $g$-orbital admixture into the
$sd$ model space is very small.

In our experiment, the state at 1616~keV was excited in the
inelastic scattering of \nuc{23}{Al} from a \nuc{9}{Be} target. In
odd-$A$ nuclei, core-coupled states are most likely excited in
inelastic scattering processes or Coulomb excitation. This is
consistent with the spin and parity assignment of $J^{\pi}=7/2^+$
for this state and the proposed structure discussed above.

\begin{figure}[h]
\epsfxsize 7.6cm \epsfbox{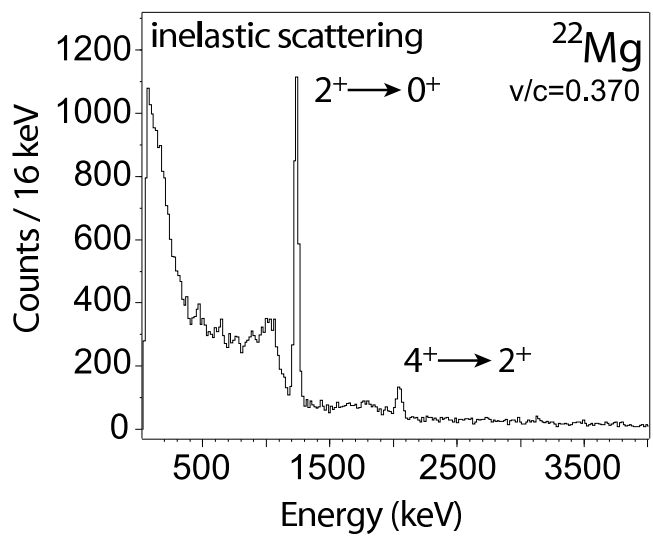}
\caption{\label{fig:mg22} Doppler-reconstructed $\gamma$-ray
spectrum obtained from the inelastic scattering of
$\nuc{9}{Be} + \nuc{22}{Mg}$ in a setting where only the
low-momentum tail of the scattered \nuc{22}{Mg} projectile
distribution entered the focal plane. The first $2^+$ and $4^+$
states are populated.}
\end{figure}

However, our analysis was restricted to scattering events with a
large momentum loss, as only the low-momentum tails of the scattered
projectiles -- including \nuc{23}{Al} -- were within the acceptance of the S800
spectrograph. To probe the nature of the states excited in these scattering
events, another constituent of the cocktail beam, \nuc{22}{Mg}, was studied
under identical conditions. In the same reaction setting where the 1616 keV
$\gamma$-ray was measured in coincidence with the low-momentum tail of the
\nuc{23}{Al} projectile distribution passing through the target, the
low-momentum tail of the \nuc{22}{Mg} projectiles within the same cocktail beam
entered the focal plane as well. The event-by-event Doppler reconstructed
$\gamma$-ray spectrum in coincidence with these inelastically scattered
\nuc{22}{Mg} projectiles is shown in Fig.~\ref{fig:mg22}. The photopeaks of the
$2^+_1 \rightarrow 0^+_1$ and $4^+_1 \rightarrow 2^+_1$ $\gamma$-ray
transitions are clearly visible. The intensities show that,
predominantly, the $2^+_1$ state of \nuc{22}{Mg} is excited in the
inelastic scattering of \nuc{22}{Mg} projectiles from the
\nuc{9}{Be} target. This is again consistent with the proposed
structure of $[\nuc{22}{Mg}(2^+_1) \otimes d_{5/2}]$ for the
1616~keV state excited in the inelastic scattering of \nuc{23}{Al}.

The proposed $(7/2^+)$ excited state in \nuc{23}{Al} was also
populated in the one-proton pickup reaction
\nuc{9}{Be}(\nuc{22}{Mg},\nuc{23}{Al}$+\gamma$)X. The \nuc{22}{Mg}
energy was 75.1~MeV/nucleon at mid-target. The potential of using
inverse-kinematics one-proton pickup reactions with fast exotic
beams for spectroscopy has recently been investigated at the NSCL
with the reaction \nuc{9}{Be}(\nuc{20}{Ne},\nuc{21}{Na}$+\gamma$)X
~\cite{gade07b}. In that case, the pattern of the observed
population of \nuc{21}{Na} single-proton states was found to be
consistent with distorted wave transfer reaction calculations and
shell-model theory. There also, a core-coupled $7/2^+$ state was
observed in \nuc{21}{Na}, with a partial cross section of 0.20(5)~mb
from an inclusive cross section of 1.85(12)~mb~\cite{gade07b}. While
unobserved feeding from higher-lying states could not be excluded,
the direct population of this state with a complex structure would
indicate the presence of higher-order processes, as for example the
pickup onto \nuc{20}{Ne} in its $2^+_1$ excited state.

For the one-proton pickup to \nuc{23}{Al}, the subject of the
present paper, an inclusive cross section of $\sigma=0.54(5)$~mb was
measured for the \nuc{9}{Be}(\nuc{22}{Mg},\nuc{23}{Al})X  reaction.
From the $\gamma$-ray intensity, a partial cross section of
$\sigma(7/2^+) \geq 0.07(2)$~mb was determined with $\sigma(7/2^+)=0.07(2)$~mb
for $\Gamma_p=0$~\footnote{We note that our experiment was not designed to
  detect protons or to separate \nuc{22}{Mg} produced by the proton decay of
  \nuc{23}{Al} from \nuc{22}{Mg} produced in projectile fragmentation and thus
  the cross section for the population of the $7/2^+$ determined from
  $\gamma$-ray spectroscopy is a lower limit if $\Gamma_p \neq 0$.}. The populations of the
core-excited $7/2^+$ states in \nuc{21}{Na}~\cite{gade07b} and
\nuc{23}{Al} (present work), at about 11\% and 14\%, respectively,
are consistent for the two measurements and suggestive of the
proposed assignment in \nuc{23}{Al}. The low-momentum tail of the
\nuc{23}{Al} one-proton pickup residue distribution (evident in
Fig.~\ref{fig:pid_pick}(b)) is also indicative of the presence
of multi-step or unbound $^8$Li final state contributions in the
reaction process. The population of other (single-proton) states in
\nuc{23}{Al}, other than the $5/2^+$ ground and $7/2^+$ excited
state could not be observed as these states decay by proton
emission~\cite{datasheets}.

\begin{figure}[h]
\epsfxsize 7.6cm \epsfbox{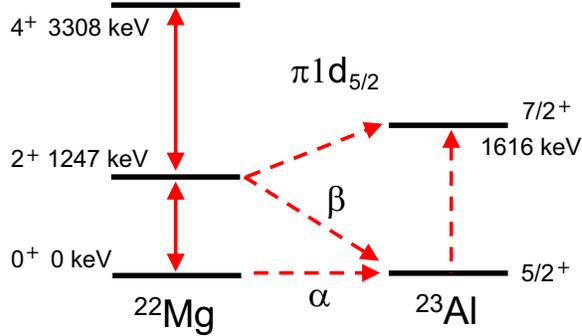} \caption{\label{fig:reaction}
Schematic of the channel coupling scheme for the coupled channels
Born approximation calculations of the $^{22}$Mg($^9$Be,
$^8$Li)$^{23}$Al($j^\pi$) reaction.}
\end{figure}

A reaction analysis of the present $\nuc{9}{Be}(\nuc{22}{Mg},
\nuc{23}{Al })X$ proton pickup data was performed within the
finite-range, post form of the coupled channels Born approximation
(CCBA) treating direct multi-step reactions via transfer and inelastic channels
using
the direct reactions code {\sc fresco} \cite{fresco}.
Single-step $1d_{5/2}$ proton transfer onto $^{22}$Mg($J^\pi$) core
states was assumed (from $^9$Be) according to the coupling scheme
shown in Figure \ref{fig:reaction}. It was also assumed (see Ref.\
\cite{gade07b}) that the final states were two-body. Thus the basic
proton pickup mechanism is computed as $^{22}$Mg($^9$Be,$^8$Li($
I^+$))$^{23}$Al($j^\pi$) leading to the $I^\pi = 1^+,2^+$ and $3^+$
states of $^8$Li at 75.1 MeV per nucleon incident energy. The
(absorptive) nuclear interactions were calculated, as in recent
nucleon knockout reaction studies, by double folding the neutron and
proton densities of $^{22}$Mg (from Hartree-Fock calculations) and
of $^9$Be (assumed a Gaussian with rms radius of 2.36 fm) with an
effective nucleon-nucleon (NN) interaction \cite{Tos04}. The
$^{22}$Mg was allowed to inelastically excite by deforming the
entrance channel distorting potential, taking a deformation length
$\delta_2 = 1.95$ fm. This corresponds to an assumed $^{22}$Mg mass
$\beta_2$ value of 0.58; this is taken from the charge $\beta_2$
value $0.58(11)$ of Ref.\ \cite{Raman}. The (light) target-like
vertices, [$^8$Li$(I^+) \otimes \Phi_j ]_{3/2^-}$, were treated as
in Ref.\ \cite{gade07b}, making use of the Variational Monte Carlo
(VMC) overlaps and spectroscopic amplitudes \cite{BobW}.

The required proton-core projectile overlaps, [$^{22}$Mg$(J^\pi)
\otimes \pi 1d_{5/2}]_{j}$, and their spectroscopic amplitudes were
guided by the USDB shell model calculations. As is noted in Fig.\
\ref{fig:reaction} there are interfering spectroscopic amplitudes
$\alpha$ and $\beta$ for population of the $^{23}$Al(gs) via the
[$^{22 }$Mg$(0^+) \otimes \pi 1d_{5/2}]_{5/2}$ and [$^{22} $Mg$(2^+)
\otimes \pi 1d_{5/2}]_{5/2}$ transfers, respectively. The $^{23
}$Al($7/2^+$) state is populated by [$^{22}$Mg$(2^+) \otimes \pi
1d_{5/2}]_{7/2}$. The USDB shell model spectroscopic factors of
these overlaps are $S$=0.33, 0.84 and 0.46, respectively. The
associated $\pi 1d_{5/2}$ single particle states were calculated in
real Woods-Saxon potential wells with radius and diffuseness
parameters $r_0=1.25$ fm and $a_0=0.7$ fm and a spin-orbit
interaction of strength 6 MeV with the same geometry parameters. The
bound $\pi 1d_{5/2}$ configurations used the physical separation
energies. The very narrow resonant $\pi 1d_{5/2}$ state, relevant to
the $7/2^+$ transition with $S_p=-244$~keV, was described
by a bound proton state with separation energy of $S_p=+5$~keV.

The calculated yields, inclusive with respect to the three
$^8$Li$(I^+)$ final state contributions, were as follows. The cross
section for direct population of the $^{23}$Al($7/2^+$) state is
0.022 mb, underproducing the measured yield of 0.07(2) mb. The
calculations were also sensitive to the relative phase of the
spectroscopic amplitudes $\alpha$ and $\beta$ that feed the
$^{23}$Al(gs), Fig.\ \ref{fig:reaction}. The shell-model
calculations for the spectroscopic amplitudes together with the
electromagnetic matrix element, predicts
that these paths interfere
destructively to give $^{23}$Al(gs) and inclusive cross sections of
0.26 and 0.28 mb, the latter to be compared with the experimental
value of 0.54(5) mb. Inelastic $5/2^+ \rightarrow 7/2^+$ (single
particle) coupling in $^{23}$Al, shown in Fig.\ \ref{fig:reaction},
was found to have negligible effect on the calculated $^{23}$Al($
7/2^+$) yield. We conclude that the {\em relative} yields of the
$^{23}$Al($5/2^+$) and $^{23}$Al($7/2^+$) are reasonably reproduced
by the model calculations. The inclusive cross section is a factor two
smaller than that measured. These lower cross sections were not
unexpected since we include only the three (lowest) $^8$Li final
states with summed spectroscopic factors of 0.97 ($p_{3/2}$) and
0.36 ($p_{1/2}$). Further consideration of strength leading to the
$^8$Li continuum is needed to assess the absolute cross sections. As
was noted above, the low-momentum tail seen in the \nuc{23}{Al}
distribution in Fig.~\ref{fig:pid_pick} suggests a significant
missing dissipative mechanism such as coupling to the $^8$Li
continuum.

In summary, the $\gamma$-decay of an excited state in \nuc{23}{Al}
has been observed for the first time in (i) \nuc{23}{Al} inelastic
scattering from \nuc{9}{Be} and (ii) in
the heavy-ion induced one-proton pickup reaction
\nuc{9}{Be}(\nuc{22}{Mg},\nuc{23}{Al}$+\gamma$)X. The corresponding
excited state at 1616(8)~keV lies 1494~keV above the proton
separation energy of \nuc{23}{Al}. The presence of the $\gamma$-ray
decay branch and the population of this state in the two reaction
mechanisms is consistent with the state being the $7/2^+$
core-excited configuration, $[\nuc{22}{Mg}(2^+_1) \otimes
d_{5/2}]_{7/2^+}$, predicted by the shell model to be the second
excited state. This picture is also consistent with the relative yields of the
ground and $7/2^+$ states obtained using multi-step, CCBA reaction
calculations. We have shown that the proton decay of this state,
which will proceed by emission of a proton from the $d_{5/2}$ orbit
to the first $2^+$ state of \nuc{22}{Mg}, is hindered by the small
$Q$-value of $Q_p=244(21)$~keV. A branching ratio of $\Gamma_{
\gamma}/\Gamma_p \sim 10$ is estimated from shell model and proton
decay calculations.

This work was supported by the National Science Foundation under
Grants No. PHY-0606007, PHY-0555366 and PHY-0653323 and by the United Kingdom
Science and Technology Facilities Council (STFC) under Grant
EP/D003628.


\end{document}